\documentclass[11pt,a4paper]{article}
\usepackage{placeins}
\usepackage[margin=2.35cm]{geometry}
\usepackage{jheppub}
\usepackage{amsmath,amssymb,bm}
\usepackage{booktabs,tabularx,array}
\usepackage{enumitem}
\usepackage{xcolor}
\usepackage{mathrsfs}
\usepackage{float}
\usepackage{caption}
\usepackage{multirow}
\setlist[itemize]{leftmargin=2em,itemsep=0.25em,topsep=0.4em}
\setlist[enumerate]{leftmargin=2.2em,itemsep=0.25em,topsep=0.4em}
\renewcommand{\arraystretch}{1.2}
\newcolumntype{Y}{>{\raggedright\arraybackslash}X}

\newcommand{\Xe}{\mathrm{Xe}}

\title{Inelastic from the Other Side: Xenon Excitation Signals in Light of the LZ High-Recoil Event}
\author[1,2,3]{Guanhua Gu}
\author[1]{Lingfeng Li}
\author[1,4]{Shao-Song Tang}
\author[5]{Yongheng Xu}
\affiliation[1]{Department of Physics and Institute of Theoretical Physics, Nanjing Normal University, Nanjing, Jiangsu 210023, China}
\affiliation[2]{Institute of Theoretical Physics, Chinese Academy of Sciences, Beijing 100190, P. R. China}
\affiliation[3]{School of Physical Sciences, University of Chinese Academy of Sciences, Beijing 100049, P. R. China}
\affiliation[4]{School of Physical Science and Technology, Xinjiang University,Urumqi 830046, China}
\affiliation[5]{Fysisk institutt, University of Oslo, Sem S{\ae}landsvei 24, N-0371, Oslo, Norway}

\emailAdd{guguanhua@itp.ac.cn}
\emailAdd{l.f.li165@gmail.com}
\emailAdd{3581951393@qq.com}
\emailAdd{yongheng.xu@fys.uio.no}

\abstract
{The LUX-ZEPLIN (LZ) experiment recently reported a single event consistent with a nuclear recoil of $E_{\rm NR}\sim248$~keV in an extended-energy search. We study the accompanying inelastic-xenon channel, $\chi+\Xe\to\chi+\Xe^{*}$, as a complementary test of dark matter (DM) interpretations of such high-recoil events. Using the non-relativistic effective-field-theory (NREFT) framework, we compute the inelastic-xenon rates for representative scenarios including accelerated DM populations and inelastic dark matter. Their signal patterns in the S1--S2 plane are reconstructed. The nuclear de-excitation shifts these signals toward the electronic-recoil band, where they may overlap with clustered isotope backgrounds. We then propose a schematic, physics-motivated background model to estimate their detectability. For many benchmarks, the inelastic-xenon rates are $\mathcal{O}(0.1)$ of the elastic rates or below, and the projected sensitivity remains weak even at $200~{\rm tonne\,yr}$ for the scenarios investigated. Our results provide additional inelastic-xenon signal predictions for various DM scenarios and motivate improved high-energy background modeling and detector calibration for future searches.
}

\begin{document}
\maketitle

\section{Introduction}
\label{sec:introduction}
Dark matter (DM) constitutes most of the matter in the universe, with evidences including the cosmic microwave background (CMB)~\cite{Planck:2018vyg,ACT:2025fju,SPT-3G:2025bzu}, galactic dynamics~\cite{Iocco:2015xga}, and gravitational lensing~\cite{Clowe:2006eq}. Its particle nature remains unknown. Direct-detection experiments search for DM scattering in terrestrial targets, with the leading sensitivities currently provided by liquid-xenon experiments~\cite{LZ:2019sgr,PandaX-4T:2021bab,LZ:2022lsv,XENON:2023cxc,PandaX:2024qfu,XENON:2025vwd}.

The recent report of an event consistent with a high-energy nuclear recoil (NR) with $E_{\rm NR}\sim248$~keV by LZ~\cite{LZ2026} motivates interpretations beyond conventional elastic spin-independent (SI) scattering of standard Galactic-halo DM. This event has motivated many dedicated studies such as~\cite{Freese:2026sga,Wu:2026nhi,Fan:2026kxx,Lou:2026idn,Su:2026rwz,Visinelli:2026kgt,Yamashita:2026ump,Yin:2026jnn,DiMauro:2026ldr,Nomura:2026qyq,Smirnov:2026aqk,Unwin:2026rdp,McCabe:2026crm,Jeesun:2026vzo,Rodd:2026tyn,Chattopadhyay:2026ryw,Du:2026guj}. At such recoil energies the momentum transfer is large and nuclear coherence is substantially reduced~\cite{Catena:2015uha}. The absence of a corresponding excess at lower recoil energies makes an interpretation within the conventional elastic SI scenario more difficult. At such a large momentum transfer, the same DM-nucleus interaction can also excite the xenon nucleus, followed by prompt electromagnetic de-excitation~\cite{Ejiri:1993ndv,Belli:1991mx,Avignone:2000ui,Bernabei:2000qn,Baudis:2013bba,XMASS-I:2014lnb,McCabe:2015eia,Vergados:2016ytt,XMASS:2018hgt,Arcadi:2019hrw,XENON:2020fgj,Dutta:2022tav,Bednyakov:2023pwi}. Such inelastic-xenon channels may provide an additional consistency test of a high-$E_{\rm NR}$ DM interpretation. Of particular interest are $^{129}\Xe$ and $^{131}\Xe$, whose lowest excited states lie at $39.6$ and $80.2$~keV, respectively~\cite{Arcadi:2019hrw,Baudis:2013bba}. To avoid ambiguity with inelastic dark matter, we refer to DM scattering that excites the xenon nucleus, DM + Xe $\to$ DM + Xe$^{*}$, as inelastic-xenon scattering throughout this work.

The accompanying inelastic-xenon signal can provide an additional test of DM scenarios that produce large $E_{\rm NR}$. However, its interpretation is more complicated than earlier phenomenological studies suggested, which did not incorporate the structured isotope backgrounds relevant to the extended S1--S2 region considered here. At higher energies, isotope-induced and activation backgrounds can form localized structures that overlap substantially with the predicted inelastic-xenon signal. Their decay and de-excitation chains, together with modified recombination and its fluctuations in liquid xenon, further broaden and shift these populations in the S1--S2 plane. The corresponding isotope abundances can also depend strongly on the calibration history and data-taking period. These effects make a reliable background prediction difficult: the information from the recent extended-energy-range LZ study~\cite{LZ2026} does not yet allow a definitive external background fit over the extended region. We therefore do not attempt a precise reinterpretation of the present data, given the limited public information. Instead, we construct a tentative signal-plus-background framework to estimate where accompanying inelastic-xenon events could become observable in the future and how they may affect the interpretation of the current observation. 

The remainder of this paper is organized as follows. In Sec.~\ref{sec:signal_model}, we introduce the physics framework of inelastic-xenon scattering. Section~\ref{sec:signal_background} describes the detector-response simulation and background modeling in an LZ-like experiment. In Sec.~\ref{sec:results}, we present results for various scenarios, including representative weakly interacting massive particle (WIMP) interactions, high-velocity DM subpopulations, and inelastic DM models. We summarize our findings in Sec.~\ref{sec:conclusion}. Further numerical results and isotope-induced background modeling are provided in the appendix.

\section{Physics of DM-Nuclei Scattering}
\label{sec:signal_model}
We consider a massive DM particle $\chi$ scattering from a xenon nucleus, including both elastic scattering, $\chi+{}^{A}\Xe\rightarrow\chi+{}^{A}\Xe$, and inelastic nuclear scattering, $\chi+{}^{A}\Xe\rightarrow\chi+{}^{A}\Xe^{*}$. The latter is particularly relevant for $^{129}\Xe$ and $^{131}\Xe$, whose first excited states lie at $E^{*}=39.6$ and $80.2$~keV, respectively~\cite{Baudis:2013bba,Arcadi:2019hrw}. We describe the DM--nucleon interaction using the non-relativistic effective-field-theory (NREFT) framework proposed in Ref.~\cite{Fan:2010gt}, with the operator basis and nuclear-response formalism developed in Refs.~\cite{Fitzpatrick:2012ix,Anand:2013yka}. The conventional SI and SD interactions correspond to $\mathcal O_1$ and $\mathcal O_4$, respectively. In this framework, the Galilean-invariant operators are built from $\bm q$, $\bm S_\chi$, $\bm S_N$, and the transverse relative velocity $\bm v^\perp$. With the convention $\bm q=\bm p_\chi'-\bm p_\chi$, it can be written at the nuclear level as
\begin{equation}
\bm v^\perp=\bm v+\frac{\bm q}{2\mu}+\frac{E^{*}}{q^2}\bm q,
\qquad
\bm v^\perp\cdot\bm q=0,
\label{eq:vperp_definition}
\end{equation}
where $\mu$ is the DM--nucleus reduced mass. For $E^{*}=0$, this expression reduces to the usual form for elastic DM--nucleus scattering. The corresponding minimum velocity is $v_{\min}=q/(2\mu)+E^{*}/q$.

The DM interaction is defined initially at the nucleon level, but nuclear excitation probes how this perturbation acts on the many-body xenon state. At leading order, the relevant nuclear operators are obtained by promoting the one-body DM-nucleon operators to operators acting on each nucleon in the nucleus. 
Here we follow the one-body density matrix (OBDM) appraoch from Ref.~\cite{Arcadi:2019hrw}, which is compatible with the NREFT framework. In this case, the nuclear transition matrix elements are combined with the DM response to obtain the scattering amplitude. Schematically,
\begin{equation} 
\left\langle |\mathcal M_{\rm eff}|^2\right\rangle = \sum_X R_X\!\left(v^{\perp\,2},q^2\right)\, W_X(q), \label{eq:general_inelastic_response} 
\end{equation} 
where $R_X$ contains the NREFT coefficients and DM kinematics, while $W_X$ is constructed from the nuclear transition matrix elements. Here, $X$ labels the different nuclear response channels; the isospin indices are suppressed for notational simplicity. The explicit decomposition can be found in Refs.~\cite{Anand:2013yka,Arcadi:2019hrw}. We use the published xenon OBDMs and response inputs of Ref.~\cite{Arcadi:2019hrw}, obtained from shell-model calculations including \texttt{Nushell@MSU}~\cite{Brown:2014bhl}, and have verified that they reproduce the published inelastic responses. From now on, we assume isoscalar interactions.

\section{Simulation of Event Responses}
\label{sec:signal_background}
We simulate the detector responses using a customized \texttt{NEST} code~\cite{Szydagis:2011tk,Szydagis:2022ikv}, following the detector-response parameters and background decomposition of Ref.~\cite{LZ2026}. The corresponding electron-recoil (ER) and NR response bands are reconstructed, while the continuum background components are simulated using the same setup to construct our background template in the S1--S2 plane.

For inelastic-xenon scatterings, we assume that the prompt electromagnetic de-excitation of $^{129,131}{\rm Xe}^{*}$ is approximated by a photon fully absorbed sufficiently close to the nuclear recoil vertex that the two deposits are reconstructed as a single event~\cite{Arcadi:2019hrw}. Because the photon produces an ER-like ionization cascade, the additional $39.6$ or $80.2$~keV energy deposition moves the inelastic population away from the ordinary NR band and toward the ER region, while retaining a response slightly different from an electron-initiated ER.\footnote{The de-excitation photon may also produce additional signal topologies, which we leave for future work.} Due to the scarcity of publicly available calibration information, we use the default \texttt{NEST} photon response, which predicts a larger photon yield than the calibrated $\beta$ response model~\cite{Szydagis:2013sih}. The impact of replacing the photon response with an electron response is examined in App.~\ref{app:electronvsphoton}. Future experimental studies, particularly searches for $^{129}{\rm Xe}$ and $^{131}{\rm Xe}$ excitations in neutron-calibration datasets, could provide more accurate signal modeling~\cite{XENON:2020fgj}.

The extended-energy region also contains important isotope-induced backgrounds, including electron capture of $^{125}{\rm I}$ and $^{127}{\rm Xe}$, and double-electron capture of $^{124}{\rm Xe}$. These components have been studied by LZ, particularly during and after DD calibration~\cite{LZ:2024wvs}, with related measurements reported by PandaX and XENON~\cite{PandaX-4T:2024fls,XENON:2022evz}. They produce localized spectral features that may overlap with the inelastic-xenon signal region and cannot be represented by a smooth extrapolation of the low-energy ER background. Mapping these backgrounds into the S1--S2 plane is complicated by atomic de-excitation cascades. For example, $^{125}{\rm I}$ decay redistributes its energy among internal-conversion and Auger electrons and low-energy X-ray photons through a multi-step cascade~\cite{Alotiby:2019fjj,Tee:2019taa}. Although the dominant transition rates are known, the complete exclusive cascade is less well constrained. Existing liquid-xenon analyses generally report reconstructed-energy spectra rather than dedicated two-dimensional S1--S2 templates for these components~\cite{LZ:2024wvs,PandaX-4T:2024fls,XENON:2022evz}. Their activities can also be time dependent: $^{133}{\rm Xe}$ and $^{125}{\rm I}$ depend strongly on neutron-calibration history, whereas $^{124}{\rm Xe}$ double-electron capture follows the natural xenon abundance. In particular, $^{125}{\rm I}$ is produced through $^{124}{\rm Xe}$ neutron activation and is subsequently removed by the getter system~\cite{LZ:2019sgr}. These complications, together with the limited public data, prevent a precise external reconstruction of the isotope backgrounds.

We therefore construct approximate two-dimensional templates for the isotope-induced backgrounds. The event-rate ranges of the $^{125}{\rm I}$ and $^{124}{\rm Xe}$ components are estimated from Ref.~\cite{LZ:2024wvs}.\footnote{We neglect the broad $^{133}{\rm Xe}$ $\beta$-decay background at $E_{\rm ER}\gtrsim80$~keV because of its short lifetime. Its contribution can be strongly reduced by excluding data taken within $\sim60$ days after DD calibration~\cite{LZ:2024wvs}.} Following the treatment motivated by Ref.~\cite{LZ2026}, we adopt a phenomenological charge-yield correction with $Q_y/Q_\beta=0.81$. With this choice, the projected yields of the leading $^{125}{\rm I}$ and $^{124}{\rm Xe}$ components show reasonable agreement with the fits of Ref.~\cite{LZ2026}, as summarized in Table~\ref{tab:isotope_comp}. Further details can be found in App.~\ref{app:isotope}.

\begin{table}[h!]
\centering

\renewcommand{\arraystretch}{1.15}
\setlength{\tabcolsep}{7pt}
\begin{tabular}{lccc}
\toprule
Process & Global yield & Predicted ROI & LZ fit~\cite{LZ2026} \\
\midrule
$^{125}\mathrm{I}$ & $2811\pm610$ & $13.4\pm2.9$ & $9.8\pm2.3$ \\
$^{124}\mathrm{Xe}$ & $784\pm112$ & $15.8\pm2.3$ & $22.6\pm5.0$ \\
\bottomrule
\end{tabular}
\caption{Projected isotope yields and ROI contributions. The global yields from Ref.~\cite{LZ:2024wvs} are multiplied by the simulated acceptances $0.0047715$ for $^{125}\mathrm I$ and $0.0201575$ for $^{124}\mathrm{Xe}$, using $Q_y/Q_\beta=0.81$. The displayed yield uncertainties propagate the global-normalization uncertainties.}
\label{tab:isotope_comp}
\end{table}

\begin{figure}
    \centering
    \includegraphics[width=0.7\linewidth]{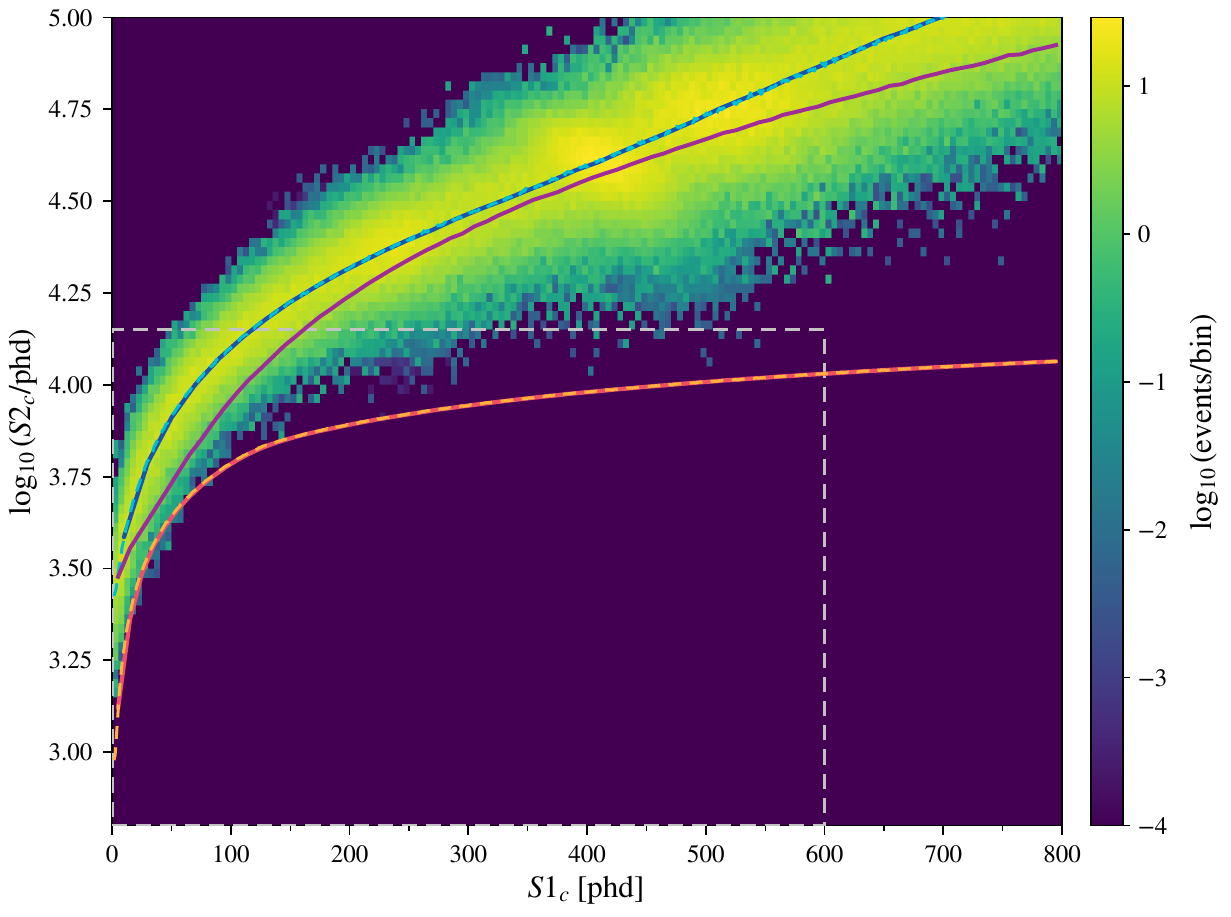}
    \caption{The background event distribution we adopt for this work in the S1--S2 plane, normalized to an exposure of 2.8 ${\rm tonne\,yr}$. We also show our ER band in solid blue curves, together with the ER band from Ref.~\cite{LZ2026} in dashed cyan curves. The NR band from our simulation (solid red curves) and Ref.~\cite{LZ2026} (dashed orange) are also shown together. In addition, we add a central line for photonic signals from our simulation as the solid purple curve slightly below the ER band. The gray box indicates the LZ region of interest.}
    \label{fig:bkg}
\end{figure}

Given the uncertainties in the isotope activities, cascade modeling, and detector response, we treat the resulting background likelihood as an empirical template rather than a precision reconstruction of the experimental background model. Its purpose is to illustrate how the high-energy background structures affect the visibility of inelastic-xenon signals in different DM scenarios. A reliable fit over the full extended-energy region would require dedicated experimental data, a more complete treatment of the isotope decays and de-excitation cascades, and improved detector-level simulations, which are beyond the scope of this work.

\section{Inelastic-Xenon Signals from Different DM Scenarios}
\label{sec:results}
We present results for several classes of dark-matter scenarios capable of producing large nuclear-recoil energies, including WIMP scattering through representative NREFT operators, inelastic dark matter (iDM), and accelerated DM populations. Our focus is on the accompanying inelastic-xenon signal associated with nuclear excitation. Although this channel need not provide the dominant sensitivity for discovery, it offers a complementary consistency test of a high-$E_{\rm NR}$ DM interpretation and may become increasingly useful with larger exposures. The study also motivates dedicated experimental analyses with improved high-energy background modeling, as well as further nuclear-theory calculations of excitation channels for xenon and other target isotopes used in direct-detection experiments. Since the LZ collaboration has conducted a statistical analysis for many elastic scattering scenarios~\cite{LZ2026}, we adopt a conservative statistical treatment and instead focus on the implications of the inelastic-xenon signal.

\subsection{WIMP Scattering in the Standard Halo Model.}
\label{ssec:vanilla_results}

\begin{figure}[h!]
    \centering
    \includegraphics[width=0.45\linewidth]{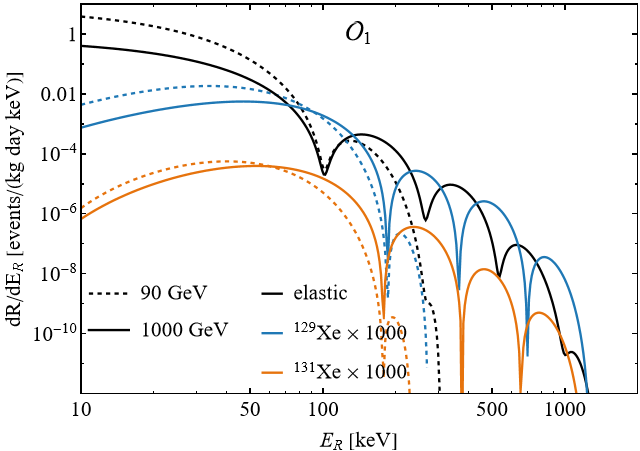}
    \includegraphics[width=0.45\linewidth]{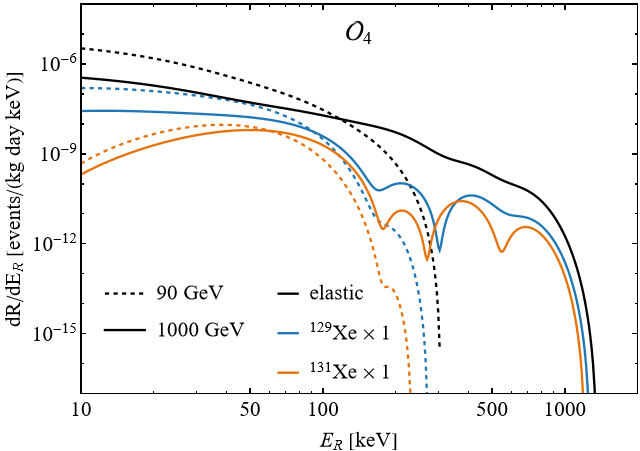}
    \includegraphics[width=0.45\linewidth]{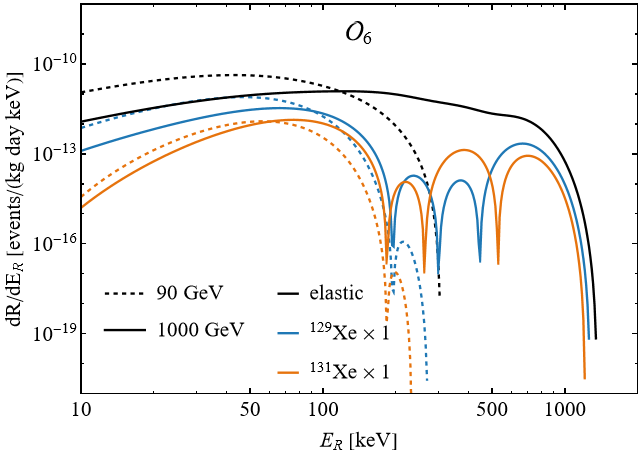}
    \includegraphics[width=0.45\linewidth]{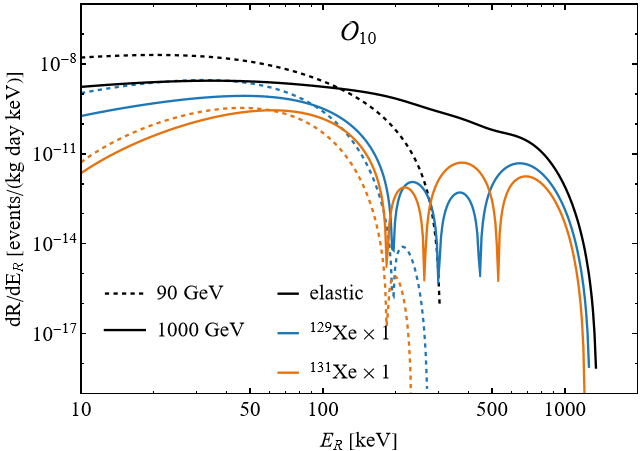}
    \caption{The expected recoil spectrum for NREFT operators for $m_{\chi}= 90$~GeV and $m_{\chi}= 1$~TeV up to the high $E_{\rm NR}$ region. The black curve stands for the total elastic rate contributed by all xenon isotopes. The cyan(orange) curve are $^{129(131)} {\rm Xe}$ inelastic-xenon rates, while solid(dashed) curves correspond to $m_\chi=$1~TeV(90~GeV) cases, respectively. Multiplicative display factors for the excitation spectra are indicated in the panel legends and must be removed when comparing physical rates. The benchmarks of $\mathcal{O}_1$ (upper-left), $\mathcal{O}_4$ (upper-right), $\mathcal{O}_6$ (lower-left), $\mathcal{O}_{10}$ (lower-right) are shown. In all plots, the NREFT cutoff scales are set to be~1~TeV.}
    \label{fig:pdsigma_NREFT}
\end{figure}

We first study the results from the WIMP case with NREFT cutoff scales set to $\mathcal{O}$(TeV). A standard halo model (SHM) with a local velocity dispersion of 220~km/s and an escape velocity cutoff of 544~km/s is used. The Earth's orbital motion is included using the monthly averaged laboratory-frame speeds following Ref.~\cite{McCabe:2013kea}. We calculate the differential event rate for each month and take the average over the twelve months. This provides an approximation to the annual mean rate, as the LZ exposure covers roughly one year. We present the representative differential event rate from NREFT operators in Fig.~\ref{fig:pdsigma_NREFT}, with inelastic-xenon rates from $^{129} {\rm Xe}$ and $^{131} {\rm Xe}$ rescaled accordingly for proper display. In the lower-left panel, $\mathcal{O}_6$ is of special interest for its preference for a high $E_{\rm NR}$, leading to high local significances in the latest LZ analysis in the form of its $\textbf{L}_{4,20}$ and partly $\textbf{L}_{10}$. \footnote{We note that the inelastic-xenon responses for $\mathcal{O}_7$, $\mathcal{O}_{13}$, and $\mathcal{O}_{14}$ may be of particular interest but require further verification. In particular, the relevant inelastic form factors for $\mathcal{O}_7$ and $\mathcal{O}_{13}$ are not directly available from public \texttt{DMFormFactor} implementations. Reliable quantitative predictions require additional nuclear-response calculations and consistency checks~\cite{Catena:2015uha,Haxton:2008zza,Haxton:2024amf}, which we leave for future investigation.}We see that with $|\boldsymbol{q}|^4$ dependence, the elastic interaction $E_{\rm NR}$ distribution peaks around 200~keV and extends to $\sim 1$~MeV when $m_\chi=$1~TeV, while the rate has a cutoff around 300~keV for the $m_\chi=90$~GeV benchmark due to inefficient energy transfer from the lighter DM particle to the $\mathcal{O}(100)$~GeV xenon nucleus. Both inelastic-xenon interactions also show a peaked structure, with the corresponding maximum reached at lower $E_{\rm NR}\sim 100$~keV, and drop sharply at higher energies. At even higher energies up to $\sim 1$~MeV, the rate shows several features with subdominant contributions to the overall rate. The results of $\mathcal{O}_{1,4,10}$ are also shown.

\begin{table}[h!]
    \centering
    \renewcommand{\arraystretch}{1.25}
    \setlength{\tabcolsep}{12pt}

    \begin{tabular}{c|ccc}
        \hline
        $m_\chi$ [GeV]
        & $\mathcal{O}_4$
        & $\mathcal{O}_6$
        & $\mathcal{O}_{10}$ \\
        \hline
        \multicolumn{4}{c}{$10\leq E_R\leq260$ keV} \\
        \hline
        90
        & $0.095\;(0.0091)$
        & $0.14\;(0.019)$
        & $0.14\;(0.017)$ \\

        300
        & $0.16\;(0.036)$
        & $0.12\;(0.038)$
        & $0.18\;(0.047)$ \\

        1000
        & $0.17\;(0.048)$
        & $0.11\;(0.042)$
        & $0.18\;(0.057)$ \\
        \hline
        \multicolumn{4}{c}{$0\leq E_R\leq2000$ keV} \\
        \hline
        90
        & $0.056\;(0.005)$
        & $0.14\;(0.020)$
        & $0.13\;(0.016)$ \\

        300
        & $0.10\;(0.022)$
        & $0.10\;(0.034)$
        & $0.17\;(0.044)$ \\

        1000
        & $0.12\;(0.031)$
        & $0.086\;(0.036)$
        & $0.16\;(0.051)$ \\

        \hline
    \end{tabular}
    
\caption{
    Ratio of the inelastic scattering rates on $^{129}$Xe and $^{131}$Xe
    relative to the elastic scattering rate for the SHM.
    Each entry is shown as $R_{129}/R_{\rm elastic}\, (R_{131}/R_{\rm elastic})$. The recoil-energy integration ranges are $10$--$260$ keV for the upper section of the table and $0$--$2000$ keV for the lower section. In all possible cases, the inelastic-xenon rates from $\mathcal{O}_1$ are always significantly lower ($<10^{-4}$ of the elastic ones) and are always ignored.
    }    
    \label{tab:SHM_inelastic_elastic_ratio}
\end{table}

\begin{figure}
    \centering
    \includegraphics[width=0.7\linewidth]{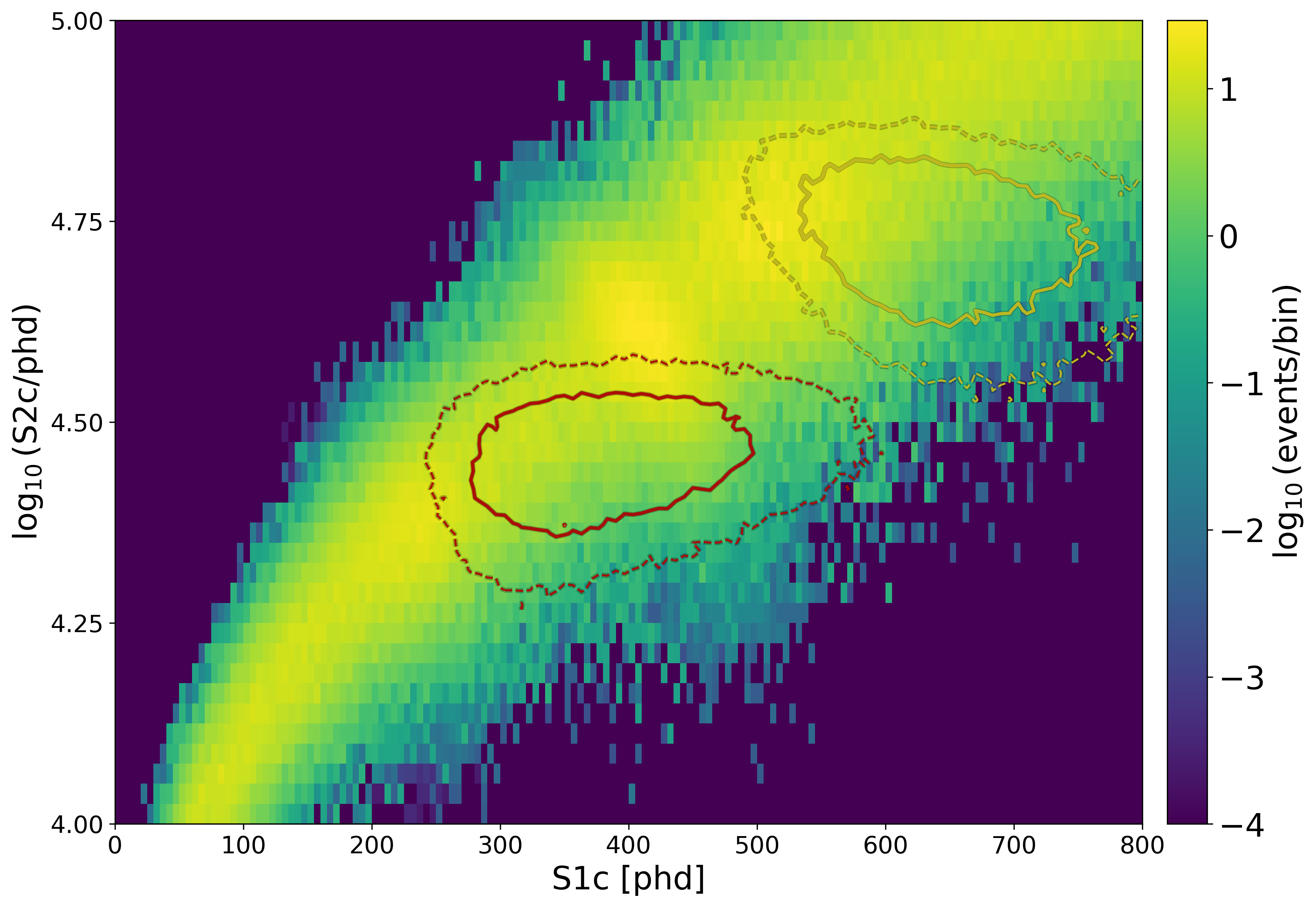}
    \caption{The zoomed-in view of inelastic-xenon scattering signal regions in the S1--S2 plane for WIMP benchmarks, together with a zoomed view of the background distribution shown in Fig.~\ref{fig:bkg} with the same exposure. Red and orange contours denote the $^{129}$Xe and $^{131}$Xe inelastic signals, respectively. For each isotope, the inner and outer contours enclose 68\% and 95\% of the simulated events, respectively. The NR contribution displaces the combined NR-plus-photon signal from the pure ER band, potentially improving its separation from ER backgrounds. However, they may overlap with the large $^{125}{\rm I}$ and $^{124}{\rm Xe}$ contributions and may be affected by the $^{133}\rm{Xe}$ contribution, which is not included in this plot due to its short lifetime.}
    \label{fig:inelastic_signal_WIMP_1}
\end{figure}

The integrated inelastic vs. elastic interaction rates for various WIMP benchmarks are listed in Table~\ref{tab:SHM_inelastic_elastic_ratio}, with contributions from the two xenon isotopes separated. Considering the finite $E_{\rm NR}$ range for current direct detection experiments, the ratios are presented for two $E_{\rm NR}$ ranges, namely $E_{\rm NR} \in [10,260]$ and [0,2000]~keV, respectively.  As shown in this table, the overall inelastic-xenon scattering rate is only of $\mathcal{O}(0.1)$ or smaller for all operators considered here, compatible with Ref.~\cite{Arcadi:2019hrw}. 
In Fig.~\ref{fig:inelastic_signal_WIMP_1}, we present the benchmark inelastic-xenon scattering of $m_\chi=$ 1~TeV with $\mathcal{O}_6$ in the S1--S2 plane. The continuous ER with isotope-induced backgrounds is also shown.

To quantify the detectability of inelastic-xenon signals, we compare the profiled likelihood including the predicted signal with the background-only distributions. Since the absolute value of the likelihood depends on statistical fluctuations of the sparse high-energy background, we normalize the likelihood shift to the spread obtained from background-only pseudo-experiments. We adopt the profiled likelihood $\mathscr{L}$ definition in~\cite{Cowan:2010js}.\footnote{We choose background yields from $^{124,127}{\rm Xe}$, $^{214} {\rm Pb}$, and $^{125}{\rm I}$ as nuisance parameters.} We then define the log-likelihood shift schematically as
\begin{equation}
\Delta \ln\mathscr{L}
\equiv -\langle\log\mathscr{L}(S+B)-\log\mathscr{L}(B)\rangle~,
\end{equation}
averaged over mock samples, where $B$ and $S+B$ stand for the background-only and signal + background assumptions, respectively. Its size is then characterized by the natural log-likelihood spread evaluated over background-only realizations ($\sigma_{\ln\mathscr{L},0}$). The ratio $\Delta\ln\mathscr{L}/\sigma_{\ln\mathscr{L},0}$ is used here only as a convenient and intuitive indicator of how strongly the predicted inelastic-xenon signal can shift the likelihood. Numerically, we find that all log-likelihood distributions are sufficiently Gaussian, so $\Delta\ln\mathscr{L}/\sigma_{\ln\mathscr{L},0}$ can be interpreted as a good indicator of the inelastic-xenon signal significance. This is especially true for the typical shift magnitude $\lesssim 1$ for scenarios investigated, and a more precise statistical interpretation becomes unnecessary. Since the signal is always additive to the background and the $\Delta\ln\mathscr{L}/\sigma_{\ln\mathscr{L},0}$ values are always negative, we take the shorthand notation of $|d|/s\equiv |\Delta\ln\mathscr{L}|/\sigma_{\ln\mathscr{L},0}$ from now on.


Given the uncertain origin of the single high $E_{\rm NR}$ event reported by LZ~\cite{LZ2026}, we do not attempt to fit the elastic recoil spectrum but instead consider two illustrative normalizations for the accompanying inelastic-xenon signals. First, we fix the expected number of $^{}{\rm Xe}$ ground-state elastic events ($N_{\rm NR}^{\rm ref}$) to be one at an exposure of $2.8~{\rm tonne\,yr}$, integrated over the recoil spectrum before detector selection. This serves as a heuristic normalization rather than a fit to the observed event. Additionally, we adopt fixed NREFT cutoff scales $\Lambda\sim\mathcal{O}({\rm TeV})$, chosen to give $N_{\rm NR}^{\rm ref}<1$, avoiding existing experimental constraints. Table~\ref{tab:stat_WIMP} shows the corresponding profiled log-likelihood shift values for an exposure of $200~{\rm tonne\,yr}$, achievable for future xenon experiments such as DARWIN~\cite{DARWIN:2016hyl} and PandaX-xT~\cite{PANDA-X:2024dlo}. Since the experimental performances of these future experiments are not yet available, we retain the current LZ detector setup for this projection. To limit background-modeling uncertainties, we restrict the analysis to S1$<800$ and S2$<10^5$. 

The small likelihood shifts shown in Table~\ref{tab:stat_WIMP} indicate that the accompanying inelastic-xenon signals have limited statistical sensitivity, even at the projected exposure of $200~{\rm tonne\,yr}$. This is mainly due to two effects: the relatively low inelastic scattering rates and the substantial overlap with the ER background, as depicted in Fig.~\ref{fig:inelastic_signal_WIMP_1}. The detector-response fluctuations smear ER events into the expected signal region. Consequently, the shifts remain small compared with the background-only likelihood fluctuations, corresponding to weak discrimination between the signal-plus-background and background-only hypotheses. 

\begin{table}[h!]
\centering
\small
\renewcommand{\arraystretch}{1.12}
\setlength{\tabcolsep}{3pt}
\begin{tabular*}{0.8\textwidth}{c cc cc cc}
\toprule
& \multicolumn{2}{c}{$\mathcal O_{4}$} & \multicolumn{2}{c}{$\mathcal O_{6}$} & \multicolumn{2}{c}{$\mathcal O_{10}$} \\
\cmidrule(lr){2-3}
\cmidrule(lr){4-5}
\cmidrule(lr){6-7}
$m_\chi$ [GeV] & $N_{\mathrm{NR}}^{\mathrm{ref}}=1$ & $\Lambda=3\,\mathrm{TeV}$ & $N_{\mathrm{NR}}^{\mathrm{ref}}=1$ & $\Lambda=3\,\mathrm{TeV}$ & $N_{\mathrm{NR}}^{\mathrm{ref}}=1$ & $\Lambda=3\,\mathrm{TeV}$ \\
\midrule
$90$ & $<0.01$ & $<0.01$ & $0.012$ & $<0.01$ & $<0.01$ & $<0.01$ \\
$1000$ & $0.041$ & $<0.01$ & $0.030$ & $<0.01$ & $0.039$ & $<0.01$ \\
\bottomrule
\end{tabular*}
\caption{Central values of $|d|/s$ for SHM WIMP benchmarks at $200$ tonne-years. For each operator, the two columns compare one ground-state event before detector selection at $2.8$ tonne-years ($N_{\mathrm{NR}}^{\mathrm{ref}}=1$) with a fixed scale $\Lambda=3$ TeV.}
\label{tab:stat_WIMP}
\end{table}

\subsection{High-velocity dark-matter subpopulations}
\label{ssec:high_velocity_subpopulation}
\begin{figure}[h!]
    \centering
    \includegraphics[width=0.48\linewidth]{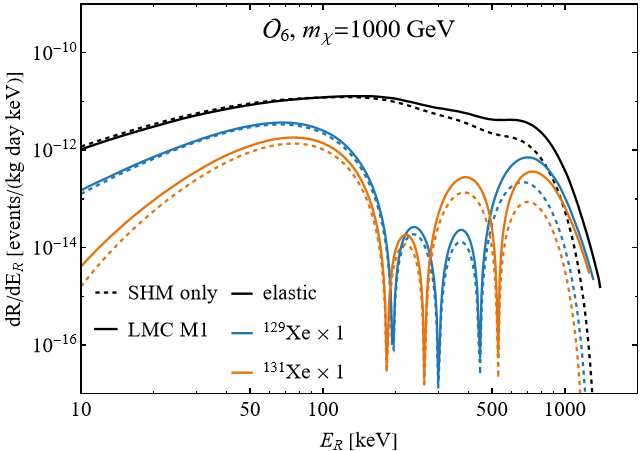}
    \includegraphics[width=0.48\linewidth]{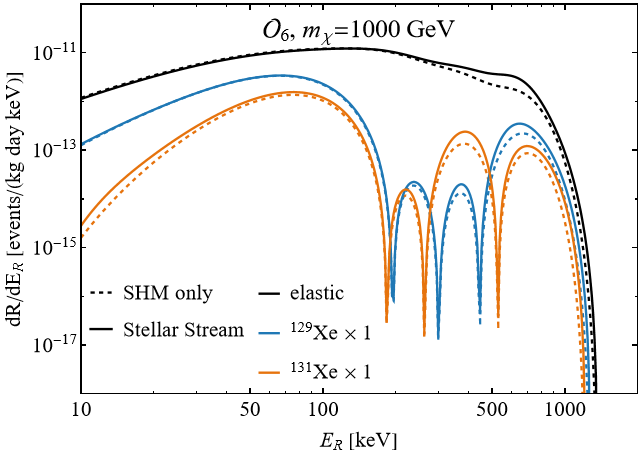}
    \includegraphics[width=0.48\linewidth]{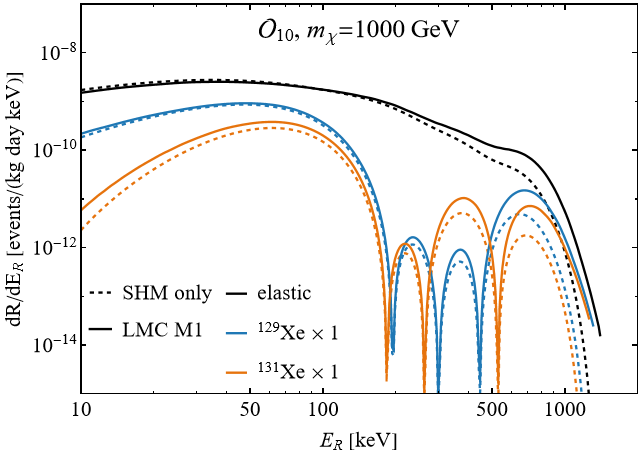}
    \includegraphics[width=0.48\linewidth]{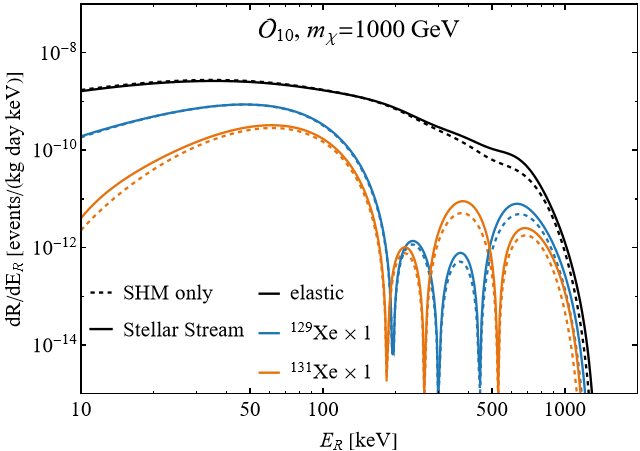}
    \caption{Similar to Fig.~\ref{fig:pdsigma_NREFT} but for accelerated DM scenarios. Here we consider the case of halo models including LMC~\cite{Besla:2019xbx} (left panels) and the stellar stream (right panels) cases. In all plots, we show corresponding SHM results in dashed curves for comparison. The top(bottom) panels correspond to $\mathcal{O}_{6(10)}$, respectively.}
    \label{fig:dsigmadE_Acc_1}
\end{figure}

A subdominant high-velocity DM population can substantially enhance the rate of large-$E_{\rm NR}$ events while contributing only a small fraction of the local DM density. We therefore consider representative accelerated Galactic subpopulations in addition to the standard halo model, using the Large Magellanic Cloud (LMC) contribution~\cite{Besla:2019xbx,Reynoso-Cordova:2024xqz} and cold stellar streams~\cite{OHare:2018trr,Buch:2019aiw} as two benchmark scenarios. 

The nearby LMC halo provides a broad high-velocity contribution that can extend beyond the nominal SHM cutoff, while a cold stream produces a narrower feature in velocity space. In the following, we use the corresponding velocity distributions of Model 1 described in Refs.~\cite{Besla:2019xbx,OHare:2018trr} and focus only on their impact on the high-recoil and accompanying inelastic-xenon signals. For the stellar stream, we model the Galactic-frame velocity distribution as a narrow component centered at the stellar stream $S_1$, making up 10\% of the local dark matter density~\cite{OHare:2018trr}.

We present the $dR/dE_{\rm NR}$ plots for WIMP benchmark cases in~Fig.~\ref{fig:dsigmadE_Acc_1}. Compared to the results from SHM, both LMC and stellar stream cases show a moderate enhancement in the scattering rate, mainly in the high-$E_{\rm NR}$ region, with the LMC having greater impact. Neither produces a large change of the major peak shape of elastic scattering. For inelastic scattering, the effects are more apparent for the $\gtrsim 300$~keV range. The corresponding likelihood results are summarized in
Table~\ref{tab:stat_LMC_S1}, using the same normalization
prescriptions as in Table~\ref{tab:stat_WIMP}.
Numerically, we find that neither of the two scenarios will significantly change the statistical conclusions compared to their SHM counterparts. 


\begin{table}[h!]
\centering
\small
\renewcommand{\arraystretch}{1.12}
\setlength{\tabcolsep}{3pt}
\begin{tabular*}{0.8\textwidth}{c cc cc cc}
\toprule
& \multicolumn{2}{c}{$\mathcal O_{4}$} & \multicolumn{2}{c}{$\mathcal O_{6}$} & \multicolumn{2}{c}{$\mathcal O_{10}$} \\
\cmidrule(lr){2-3}
\cmidrule(lr){4-5}
\cmidrule(lr){6-7}
$m_\chi$ [GeV] & $N_{\mathrm{NR}}^{\mathrm{ref}}=1$ & $\Lambda=3\,\mathrm{TeV}$ & $N_{\mathrm{NR}}^{\mathrm{ref}}=1$ & $\Lambda=3\,\mathrm{TeV}$ & $N_{\mathrm{NR}}^{\mathrm{ref}}=1$ & $\Lambda=3\,\mathrm{TeV}$ \\
\midrule
\multicolumn{7}{c}{LMC} \\
\midrule
$90$ & $0.010$ & $<0.01$ & $0.026$ & $<0.01$ & $0.017$ & $<0.01$ \\
$1000$ & $0.057$ & $<0.01$ & $0.029$ & $<0.01$ & $0.045$ & $<0.01$ \\
\midrule
\multicolumn{7}{c}{$90\%$ SHM $+\,10\%$ Stellar Stream} \\
\midrule
$90$ & $<0.01$ & $<0.01$ & $0.016$ & $<0.01$ & $0.012$ & $<0.01$ \\
$1000$ & $0.046$ & $<0.01$ & $0.031$ & $<0.01$ & $0.041$ & $<0.01$ \\
\bottomrule
\end{tabular*}
\caption{Central values of $|d|/s$, similar to Table~\ref{tab:stat_WIMP} but for the LMC and $90\%$ SHM $+\,10\%$ stellar stream scenarios.} 
\label{tab:stat_LMC_S1}
\end{table}

\subsection{Inelastic dark matter}
\label{ssec:idm}
\begin{figure}[h!]
    \centering
    \includegraphics[width=0.48\linewidth]{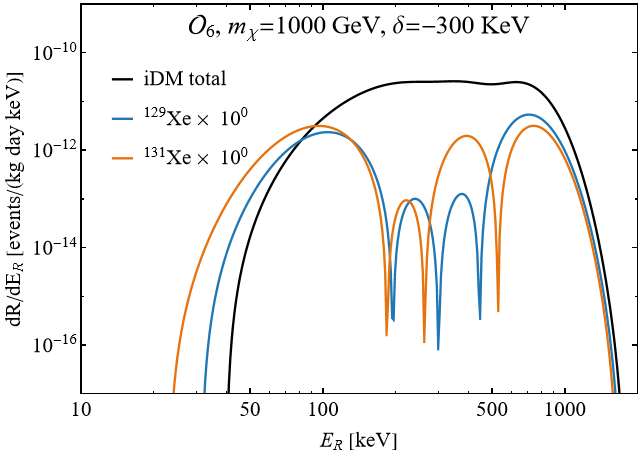}
    \includegraphics[width=0.48\linewidth]{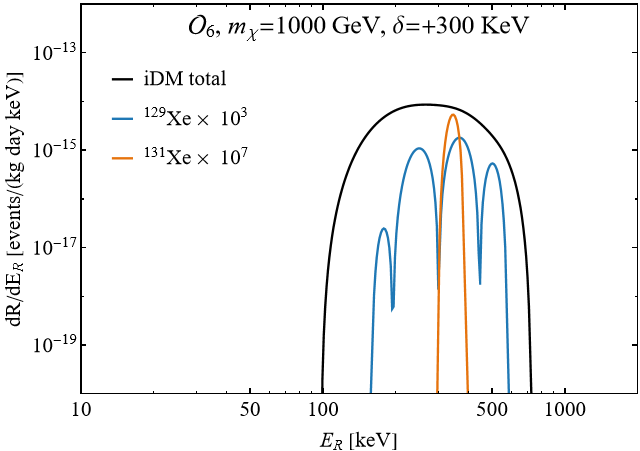}
    \includegraphics[width=0.48\linewidth]{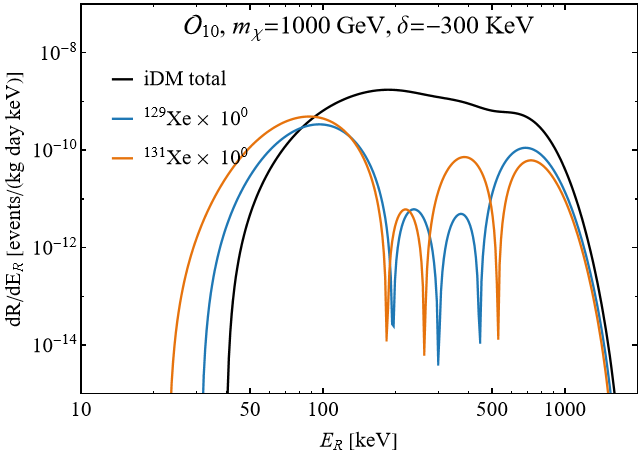}
    \includegraphics[width=0.48\linewidth]{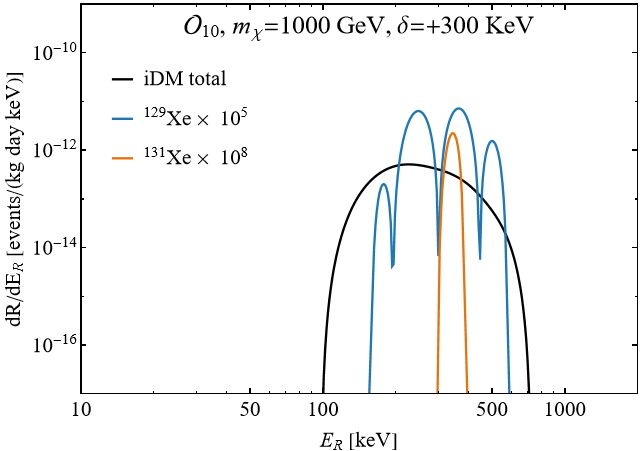}\\
    \caption{Similar to Fig.~\ref{fig:pdsigma_NREFT}, but for different iDM benchmarks with $\delta=-300$~keV (left) and $\delta=+300$~keV (right). The upper and lower panels show the results for $\mathcal{O}_6$ and $\mathcal{O}_{10}$, respectively. Both benchmarks assume $m_\chi=1$~TeV. 
    }
    \label{fig:dsigmadE_iDM_1}
\end{figure}
Inelastic dark matter (iDM) provides another simple mechanism for enhancing the high-$E_{\rm NR}$ population. We consider scattering between two nearby dark-sector states, $\chi_1 N\rightarrow\chi_2 N^{(\ast)}$, with mass splitting $\delta\equiv m_{\chi_2}-m_{\chi_1}$~\cite{Tucker-Smith:2001myb,Tucker-Smith:2004mxa,Chang:2008gd,Graham:2010ca}. The minimum incoming velocity required for a nuclear recoil $E_{\rm NR}$ is
\begin{equation}
v_{\min}(E_{\rm NR})=\frac{1}{\sqrt{2m_N E_{\rm NR}}}\left|\frac{m_NE_{\rm NR}}{\mu}+\delta+E^*\right|~,
\end{equation}
where $E^*$ is the nuclear excitation energy. For exothermic scattering with $\delta+E^*<0$, the released energy can be converted into nuclear recoil energy, shifting the spectrum toward larger $E_{\rm NR}$~\cite{Tucker-Smith:2001myb,Tucker-Smith:2004mxa,Chang:2008gd,Graham:2010ca}. 
A splitting of order $\mathcal{O}(100)\,{\rm keV}$ can therefore naturally populate the high-recoil region without requiring an anomalously large incoming velocity. In this sense, the scenario is inelastic in both the SM and DM sectors when considering $^{129,131}{\rm Xe}^\ast$ or ``doubly-inelastic". This possibility has also been considered recently in connection with high-energy xenon recoil events~\cite{An:2025bby}.

\begin{figure}[h!]
    \centering
    \includegraphics[width=1\linewidth]{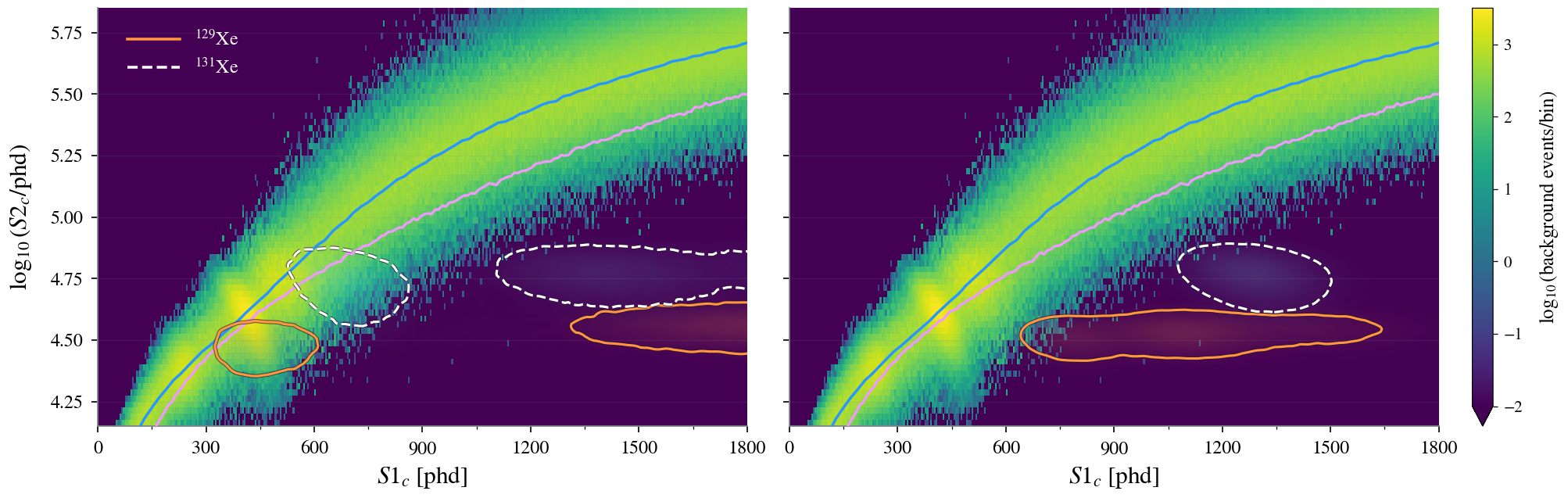}

    \caption{Signal distributions for $\mathcal{O}_6$ with $m_\chi=1$ TeV and
    $\delta=-300$ keV (left) or $+300$ keV (right). Orange solid and white dashed contours
    correspond to $^{129}$Xe and $^{131}$Xe, respectively, and contain
    approximately $90\%$ of each isotope's simulated events. Faint fills indicate independently peak-normalized signal
    densities. Blue and light-violet curves show the mean $\beta$-ER and $\gamma$-ER responses. }
    \label{fig:signal_iDM_1}
\end{figure}

\begin{table}[h!]
    \centering
    \renewcommand{\arraystretch}{1.25}
    \setlength{\tabcolsep}{12pt}

    \begin{tabular}{c|ccc}
        \hline
        $\delta$ [keV]
        & $\mathcal{O}_4$
        & $\mathcal{O}_6$
        & $\mathcal{O}_{10}$ \\
        \hline
        \multicolumn{4}{c}{$10\leq E_R\leq260$ keV} \\
        \hline

        $+300$
        & $0.0005\;(<10^{-4})$
        & $<10^{-4}\;(<10^{-4})$
        & $<10^{-4}\;(<10^{-4})$  \\
        
        $+50$
        & $0.11\;(0.024)$
        & $0.077\;(0.022)$
        & $0.11\;(0.028)$ \\
        
        $-50$
        & $0.30\;(0.12)$
        & $0.14\;(0.069)$
        & $0.24\;(0.11)$ \\

        $-300$
        & $0.15\;(0.24)$
        & $0.053\;(0.069)$
        & $0.093\;(0.13)$ \\
        \hline
        \multicolumn{4}{c}{$0\leq E_R\leq2000$ keV} \\
        \hline

        $+300$
        & $0.001\;(<10^{-4})$
        & $<10^{-4}\;(<10^{-4})$
        & $<10^{-4}\;(<10^{-4})$  \\
        
        $+50$
        & $0.096\;(0.022)$
        & $0.061\;(0.021)$
        & $0.10\;(0.026)$ \\
        
        $-50$
        & $0.35\;(0.11)$
        & $0.10\;(0.054)$
        & $0.21\;(0.096)$ \\

        $-300$
        & $0.13\;(0.17)$
        & $0.11\;(0.080)$
        & $0.088\;(0.089)$ \\
        \hline
    \end{tabular}
    \caption{Ratio of the inelastic scattering rates on $^{129}$Xe and $^{131}$Xe relative to the elastic scattering rate for inelastic dark matter with $m_\chi = 1000~\mathrm{GeV}$ in the SHM. Each entry is shown as $R_{129}/R_{\rm elastic}\, (R_{131}/R_{\rm elastic})$. The recoil-energy integration ranges are $10$--$260$ keV for the upper section of the table and $0$--$2000$ keV for the lower section.
    }
    \label{tab:SHM_iDM_inelastic_elastic_ratio}
\end{table}

For the exothermic benchmark $\delta=-300$~keV, the recoil spectrum develops a double-peak structure, as shown in the left panel of Fig.~\ref{fig:dsigmadE_iDM_1}. The two contributions are concentrated at $E_{\rm NR}\lesssim200$~keV and $E_{\rm NR}\gtrsim500$~keV, respectively. The corresponding inelastic-xenon signals form two separated populations in the S1--S2 plane, as shown in the left panel of Fig.~\ref{fig:signal_iDM_1}. In particular, the higher-energy population largely falls beyond the S1$<800$ selection, substantially reducing the observable signal and its statistical sensitivity.\footnote{Extending the signal acceptance to even higher S1 may increase the significance. However, the background estimation becomes highly challenging in the extended region. No quantitative conclusion can be drawn without further experimental inputs.}

For the endothermic benchmark $\delta=+300$~keV, shown in the right panels of Figs.~\ref{fig:dsigmadE_iDM_1} and \ref{fig:signal_iDM_1}, the large kinetic-energy threshold produces a narrow recoil spectrum around $E_{\rm NR}\sim300$~keV, potentially relevant to the high-energy LZ event~\cite{LZ2026}. In this case, the scattering rate is suppressed due to the high DM velocity threshold and is thus highly sensitive to the high-velocity tail of the DM distribution. The accompanying inelastic-xenon rate is further suppressed, making it particularly difficult to observe and sensitive to modeling uncertainties. 
The integrated inelastic-to-elastic rate ratios for iDM benchmarks are summarized in Table~\ref{tab:SHM_iDM_inelastic_elastic_ratio}, similar to Table~\ref{tab:SHM_inelastic_elastic_ratio}.


\begin{table}[h!]
\centering
\small
\renewcommand{\arraystretch}{1.12}
\setlength{\tabcolsep}{3pt}
\begin{tabular}{cc cc cc cc}
\toprule
& & \multicolumn{2}{c}{$\mathcal O_{4}$} & \multicolumn{2}{c}{$\mathcal O_{6}$} & \multicolumn{2}{c}{$\mathcal O_{10}$} \\
\cmidrule(lr){3-4}
\cmidrule(lr){5-6}
\cmidrule(lr){7-8}
$m_\chi$ [GeV] & $\delta$ [keV] & $N_{\mathrm{NR}}^{\mathrm{ref}}=1$ & $\Lambda=3\,\mathrm{TeV}$ & $N_{\mathrm{NR}}^{\mathrm{ref}}=1$ & $\Lambda=3\,\mathrm{TeV}$ & $N_{\mathrm{NR}}^{\mathrm{ref}}=1$ & $\Lambda=3\,\mathrm{TeV}$ \\
\midrule
\multirow{6}{*}{$90$} & $-300$ & $0.27$ & $0.12$ & $0.12$ & $<0.01$ & $0.16$ & $<0.01$ \\
 & $-200$ & $0.17$ & $0.096$ & $0.10$ & $<0.01$ & $0.13$ & $<0.01$ \\
 & $-50$ & $0.018$ & $0.017$ & $0.028$ & $<0.01$ & $0.025$ & $<0.01$ \\
 & $+50$ & $<0.01$ & $<0.01$ & $<0.01$ & $<0.01$ & $<0.01$ & $<0.01$ \\
 & $+200$ & $<0.01$ & $<0.01$ & $<0.01$ & $<0.01$ & $<0.01$ & $<0.01$ \\
 & $+300$ & $<0.01$ & $<0.01$ & $<0.01$ & $<0.01$ & $<0.01$ & $<0.01$ \\
\midrule
\multirow{6}{*}{$1000$} & $-300$ & $0.29$ & $0.012$ & $0.020$ & $<0.01$ & $0.059$ & $<0.01$ \\
 & $-200$ & $0.28$ & $0.016$ & $0.038$ & $<0.01$ & $0.092$ & $<0.01$ \\
 & $-50$ & $0.076$ & $<0.01$ & $0.040$ & $<0.01$ & $0.063$ & $<0.01$ \\
 & $+50$ & $0.032$ & $<0.01$ & $0.022$ & $<0.01$ & $0.026$ & $<0.01$ \\
 & $+200$ & $0.050$ & $<0.01$ & $<0.01$ & $<0.01$ & $<0.01$ & $<0.01$ \\
 & $+300$ & $<0.01$ & $<0.01$ & $<0.01$ & $<0.01$ & $<0.01$ & $<0.01$ \\
\bottomrule
\end{tabular}
\caption{Central values of $|d|/s$,similar to Table~\ref{tab:stat_WIMP} but for SHM iDM benchmarks with nonzero mass splittings.} 
\label{tab:stat_iDM}
\end{table}

We then perform statistical analyses similar to those described in Sec.~\ref{ssec:vanilla_results} and~\ref{ssec:high_velocity_subpopulation} to obtain $|d|/s$ as empirical likelihood diagnostics. Table~\ref{tab:stat_iDM} lists the projected $d/s$ values at $200$~tonne-year for the nominal $90$~GeV and $1$~TeV mass groups and mass splittings $\delta=\pm50$, $\pm200$, and $\pm300$~keV, respectively. \footnote{For the $m_{\chi}=$90~GeV, $\delta=+200(300)$~keV benchmarks, the minimal DM velocity required to have $\chi_1 +{\rm Xe}\to \chi_2 +{\rm Xe}^{(\ast)}$ scatterings will be greater than 800 km/s. Consequently, all scattering processes, with or without xenon nuclear excitations, will not happen according to SHM. The $N_{\rm NR}^{\rm ref}=1$ normalization is actually not well-defined in this case.} We note that the $|d|/s$ values, as an indicator of significance, show a non-trivial pattern when $\delta< +50$~keV, as this quantity relates to both inelastic xenon rates and the morphology of the signal region. 

\section{Conclusion}
\label{sec:conclusion}
The recent report of an event with $E_{\rm NR}\sim248$~keV by the LZ collaboration~\cite{LZ2026} has motivated a wide range of DM scenarios capable of producing such high recoil energies. We focus on the accompanying signal from inelastic excitation of the xenon nucleus, $\chi+{}^{A}\Xe\rightarrow\chi+{}^{A}\Xe^{*}$, followed by prompt de-excitation. At these recoil energies, excitation of the low-lying states of $^{129}\Xe$ and $^{131}\Xe$ is kinematically accessible, providing an additional channel to test DM interpretations. In this work, we have studied inelastic-xenon scattering using the NREFT framework and published nuclear-transition inputs, obtaining the corresponding differential scattering rates. We have also simulated their two-dimensional S1--S2 responses for representative WIMP interactions, accelerated DM subpopulations, and iDM scenarios using \texttt{NEST} with detector parameters provided by the LZ collaboration. These predictions, particularly for iDM and accelerated DM populations, extend previous inelastic-xenon studies and constitute one of the main outcomes of this work. In addition, we constructed a schematic, empirically motivated background model for the high-energy region, paying particular attention to the structured and time-dependent contributions from $^{125}{\rm I}$ and $^{124}{\rm Xe}$.


We find that most scenarios tested here produce relatively small inelastic-xenon rates, typically $\lesssim\mathcal{O}(0.1)$ of the corresponding elastic rates. Moreover, the prompt electromagnetic de-excitation shifts the signal closer to the ER band, where isotope-induced background clusters can overlap substantially with the predicted inelastic-xenon population. Our results indicate that reliable searches for these signals require a good understanding of the isotope-induced backgrounds and the smearing of the ER band. The large ER background and its (non-Gaussian) detector-response fluctuations considerably reduce the sensitivity to inelastic-xenon signals, even with an exposure of $200~\mathrm{tonne\,yr}$.

We emphasize that our current analysis is exploratory and relies on several approximations, limiting its quantitative precision. Firstly, the inelastic-xenon detector response is modeled using simplified photon- and electron-induced energy deposits, without a complete treatment of the nuclear and atomic de-excitation cascades. Although we adopt the NEST parameters reported by LZ, this does not guarantee an accurate description of the detector response throughout the extended S1--S2 region. Secondly, the isotope backgrounds introduce further uncertainties, particularly from the approximate atomic-cascade modeling and the phenomenological charge-yield correction parameter. Current public data are insufficient to fully calibrate these effects or construct a reliable experimental background likelihood. Our projected sensitivities should therefore be regarded as indicative rather than precise experimental forecasts. Nevertheless, the qualitative picture of accompanying inelastic-xenon signals, their differential $d R /d E_{\rm NR}$ distributions across DM scenarios, and their potential overlap with structured isotope backgrounds remains informative. This work provides a starting framework for more detailed experimental and phenomenological studies.

Looking ahead, larger exposures alone may not substantially improve the sensitivity for the scenarios studied without better control of the high-recoil-energy backgrounds. Improved measurements and simulations of atomic de-excitation, isotope activities, and detector charge and light yields would help constrain the background populations in the S1--S2 plane. Meanwhile, improved nuclear-transition calculations, particularly for additional NREFT operators and other target nuclei, would extend the range of testable scenarios. Together with future high-energy search data covering a wider S1--S2 region, these developments could make inelastic nuclear excitation a more useful complementary probe of DM interactions, independently of the interpretation of the present LZ event.


\section*{Acknowledgment}
We thank Chih-Ting Lu, Liang Liang Su, Lei Wu, and Wen-Na Yang for helpful discussions. We are also grateful to Wen-Na Yang for assistance with setting up the OBDM codes. Y. Xu has received funding from the European Union's Horizon Europe research and innovation programme under the Marie Skłodowska-Curie grant agreement No. 101126636. G. Gu and S. Tang are supported by the National Natural Science Foundation of China (Grant No. 12335005).

\bibliographystyle{jhep}
\bibliography{ref}

\appendix

\section{Simplified Isotope Cascade Models}
\label{app:isotope}
\subsection{Model of $^{125} {\rm I}$ Background}
\label{sapp:I125}
The $^{125}{\rm I}$ electron-capture background contains both an atomic vacancy and the excited nuclear state $^{125}{\rm Te}^\ast$ of 35.5~keV. A complete event-by-event atomic cascade is complicated and is not uniquely fixed by the available inclusive electron and X-ray measurements, especially in liquid Xenon experiments. We therefore use a simplified cascade model intended only to retain the leading photon--electron structure relevant for the S1--S2 response. In particular, we treat the relaxation of the initial atomic vacancy and the subsequent $^{125}{\rm Te}^{*}$ de-excitation as two independent steps. This approximation is motivated by the fact that atomic relaxation occurs much faster than the $1.48$~ns lifetime of the nuclear excited state~\cite{Alotiby:2019fjj}.

The initial electron capture occurs predominantly from the $K$, $L$, and $M$ shells, whose branching fractions are well measured. The corresponding branches are summarized in Table~\ref{tab:i125_ec}. 
The last column lists only the atomic-vacancy energy $E_{X^{-1}}$ and does not include the $35.5$~keV nuclear excitation. Assuming full energy deposition, the total visible energies for the listed $K$-, $L$-, and $M$-capture branches are therefore $67.3$, $40.4$, and $36.5$~keV, respectively, using the rounded energies in the table.

\begin{table}[h!]
\centering
\begin{tabular}{cccc}
\hline
EC shell & Branching fraction & State after EC & $E_{X^{-1}}$\\
\hline
$K$ & $80.11\%$ & $^{125}{\rm Te}^{*}+K^{-1}$ & $31.8$~keV \\
$L$ & $15.61\%$ & $^{125}{\rm Te}^{*}+L^{-1}$ & $4.9$~keV \\
$M$ & $3.49\%$ & $^{125}{\rm Te}^{*}+M^{-1}$ & $1.0$~keV \\
\hline
\end{tabular}
\caption{Initial $^{125}{\rm I}$ electron-capture branches used in the toy cascade. After capture, the daughter contains an excited $^{125}{\rm Te}^{*}$ nucleus together with an atomic vacancy of a particular shell noted as $X^{-1}$.}
\label{tab:i125_ec}
\end{table}

We next assign a small number of representative fates to each atomic vacancy, as summarized in Table~\ref{tab:i125_vacancy}. For a $K$ vacancy, the measured Te fluorescence yield is approximately $\omega_K\simeq0.85$--$0.88$~\cite{Alotiby:2019fjj}. We therefore use $f_K=0.88$ as the tentative probability for emission of a representative $K$ X-ray with $\sim32$~keV~\cite{Bowe1952}. The remaining non-radiative component is represented by a single KLL Auger electron with $E_{\rm KLL}\simeq22.7$~keV, with the detailed Auger-electron spectrum simplified in our treatment. The measured $K$-X-ray and KLL spectra contain several nearby lines, but resolving this fine structure is unnecessary for the present detector-level study~\cite{Alotiby:2019fjj}. For an $L$ vacancy, the microscopic fluorescence branching in liquid xenon is less certain for our purposes. The fluorescence probability associated with an $L^{-1}$ vacancy is typically small, $f_L\lesssim0.1$~\cite{XrayDBXe}. Here we take $f_L=0.1$ as a schematic benchmark and represent the radiative branch by an $L$ X-ray with $E_L\simeq3.8$~keV. For the much smaller $M$-shell contribution, we simply take $f_M=f_L$ as the $ad~hoc$ $M$-shell photon rate for simplicity, which does not significantly contradict the X-ray measurement~\cite{XrayDBXe}.

\begin{table}[h!]
\centering
\begin{tabular}{cccc}
\hline
Vacancy  & Probability & Explicit particle \\
\hline
$K^{-1}$ &  $f_K\simeq0.88$ & $\gamma_K$($\sim$32 keV) + photon-like\\
          & $1-f_K$ & $e_{\rm KLL}$($\sim$23 keV) + photon-like \\
$L^{-1}$  & $f_L\simeq 0.1$ & $\gamma_L$($\sim$3.8 keV)+ photon-like \\
          & $1-f_L$ &  photon-like \\
$M^{-1}$  & $f_M\simeq 0.1$ & $\gamma_M$($\sim$1.0 keV) \\
          & $1-f_M$ &  photon-like \\
\hline
\end{tabular}
\caption{Simplified atomic-vacancy relaxation treatment. Any vacancy energy not carried by the explicitly generated photon or Auger electron is deposited locally as a photon-like component, in which 80\% of the energy is emitted in the form of photons. The parameters $f_L$ and $f_M$ should be regarded as effective cascade parameters rather than microscopic fluorescence yields.}
\label{tab:i125_vacancy}
\end{table}

The $35.5$~keV nuclear excitation is treated separately. The direct photon branch is well measured and accounts for $6.68\%$  of monochromatic decays. The dominant internal-conversion channel is K-shell conversion, with a branching fraction close to $80\%$, producing a conversion electron with $E_{\rm CE}\simeq3.7$~keV together with a new $K^{-1}$ vacancy, as suggested by the $^{125}{\rm Te}^*$ decay measurement~\cite{Bowe1952}. This vacancy is then passed through exactly the same $K$-vacancy de-excitation as in Table~\ref{tab:i125_vacancy}. The remaining internal-conversion channels, dominated by the $L$ and $M$ shells, account for approximately $13.3\%$ of the nuclear de-excitations. To avoid introducing additional poorly constrained cascade branches, we treat their full energy as a local photon-like deposition. The resulting nuclear branches are summarized in Table~\ref{tab:i125_te}.

\begin{table}[h!]
\centering
\begin{tabular}{ccc}
\hline
Branch & Probability & Toy-model final state \\
\hline
direct $\gamma$ & $6.68\%$ & $\gamma_{\rm Te^\ast}(35.5~{\rm keV})$ \\
$K$-IC & $\simeq80\%$ & $e_{\rm CE}(\sim 3.7~{\rm keV})+K^{-1}$ \\
Outer-shell IC & $\simeq13\%$ & photon-like \\
\hline
\end{tabular}
\caption{Simplified treatment of the $35.5$~keV $^{125}{\rm Te}^{*}$ nuclear de-excitation. IC and CE denote internal conversion and conversion electron, respectively.}
\label{tab:i125_te}
\end{table}

After each explicitly generated X-ray, Auger electron, or conversion electron, the remaining energy required by energy conservation is assigned to a local photon-like component, with most energy emitted in photons. In practice, we found that the ratio $E_{\rm photon}/E_{\rm electron} = 0.8/0.2$ works with the latest LZ data and is applied to all cascade branches as the final ``energy dump". This is an over-simplified approximation to the unresolved low-energy cascades in liquid xenon. Moreover, we expect some degeneracies between such cascade uncertainty and the phenomenological charge-yield correction introduced in Sec.~\ref{sec:signal_background}. Changing $f_L$ or $f_M$ changes the fraction of the decay energy initially assigned to photons and electrons, while the subsequent $Q_y/Q_\beta$ prescription redistributes the detector-level charge and light yields. These two ingredients should therefore not be interpreted as independent precision measurements, while the best solution remains to be calibrated by data.

\subsection{Model of $^{124} {\rm Xe}$ Background}
\label{sapp:Xe124}
For $^{124}{\rm Xe}$ double-electron capture, we start from the shell-combination branching fractions and reconstructed energies reported in Ref.~\cite{LZ:2024wvs}. That analysis is primarily formulated in terms of the total reconstructed energy of single-site events, whereas our purpose requires an approximate decomposition into photon- and electron-like energy deposits in order to generate the S1--S2 response. We therefore keep the published capture-mode branching fractions and total visible energies, but further resolve the associated atomic vacancies using the same simplified treatment introduced above.

Similar to the approach in App.~\ref{sapp:I125}, we treat the de-excitation of each vacancy separately as listed in Table~\ref{tab:xe124_vacancy}. 
For example, each $K^{-1}$ vacancy produces a representative $\sim30$~keV X-ray with probability $f_K\simeq0.90$, with any residual energy assigned to the local photon-like component. With probability $1-f_K\simeq0.10$, it instead produces a representative $\sim23$~keV KLL Auger electron, followed by the residual-energy deposit. For an $L^{-1}$ vacancy, we use a representative $\sim4$~keV X-ray with probability $f_L\simeq0.06$~\cite{XrayDBXe}. The remaining energy is treated as a local photon-like deposit. For $M^{-1}$ and other vacancies, we retain only the local photon-like contribution, as summarized in Table~\ref{tab:xe124_vacancy}.

\begin{table}[t]
\centering
\begin{tabular}{cccc}
\hline
Vacancy  & Probability & Explicit particle \\
\hline
$K^{-1}$ &  $f_K\simeq0.9$ & $\gamma_K$($\sim$30 keV) + photon-like\\
          & $1-f_K$ & $e_{\rm KLL}$($\sim$23 keV) + photon-like \\
$L^{-1}$  & $f_L\simeq 0.06$ & $\gamma_L$($\sim$4 keV)+ photon-like \\
          & $1-f_L$ &  photon-like \\
$M^{-1}$ and others  & $1-f_M \sim 1$ &  photon-like \\
\hline
\end{tabular}
\caption{Simplified vacancy treatment used for the $^{124}{\rm Xe}$ double-electron-capture background.}
\label{tab:xe124_vacancy}
\end{table}

Inner-shell capture and the associated Auger and internal-conversion cascades can have a different charge yield from ordinary $\beta$-induced ER events. Motivated by the enhanced recombination observed for xenon electron-capture populations~\cite{LZ:2025hud,Xu:2025ksk} and emphasized in the recent extended-energy LZ analysis~\cite{LZ2026}, we introduce a phenomenological relative charge-yield parameter $Q_y/Q_\beta=0.81$. 
The setup preserves the total secondary electrons + photons while shifting the event toward larger light yield and smaller charge yield. The modified electron and photon yields are then converted to S1 and S2 using the corresponding LZ $g_1$ and $g_2$ parameters. The correction therefore mainly changes the location of each isotope population in the S1--S2 plane and does not alter its total normalization.

\section{Further Numerical Results for Various DM Scenarios}
\label{app:number}
In this section, we present the inelastic-xenon scattering rate details for various NREFT operators and DM scenarios.

\begin{table}[htpb]
    \centering
    \renewcommand{\arraystretch}{1.25}
    \setlength{\tabcolsep}{12pt}

    \begin{tabular}{c|ccc}
        \hline
        Model
        & $\mathcal{O}_4$
        & $\mathcal{O}_6$
        & $\mathcal{O}_{10}$ \\
        \hline
        \multicolumn{4}{c}{$10\leq E_R\leq260$ keV} \\
        \hline
        
        S1
        & $0.18\;(0.059)$
        & $0.11\;(0.048)$
        & $0.19\;(0.067)$ \\
        
        LMC M1
        & $0.20\;(0.071)$
        & $0.12\;(0.053)$
        & $0.19\;(0.077)$ \\
        \hline
        \multicolumn{4}{c}{$0\leq E_R\leq2000$ keV} \\
        \hline

        S1
        & $0.13\;(0.038)$
        & $0.11\;(0.048)$
        & $0.18\;(0.065)$ \\
        
        LMC M1
        & $0.15\;(0.046)$
        & $0.12\;(0.053)$
        & $0.19\;(0.075)$ \\

        
        \hline
    \end{tabular}
        \caption{
    Ratio of the inelastic scattering rates on $^{129}$Xe and $^{131}$Xe
    relative to the elastic scattering rate for different dark-matter
    velocity distributions at $m_\chi=1~\mathrm{TeV}$.
    Each entry is shown as
    $R_{129}/R_{\rm elastic}\,
    (R_{131}/R_{\rm elastic})$.
    }
    \label{tab:LMC_S1_inelastic_elastic_ratio}
\end{table}

\begin{figure}[p]
    \centering

    \includegraphics[
        width=0.48\linewidth,
        height=0.24\textheight,
        keepaspectratio
    ]{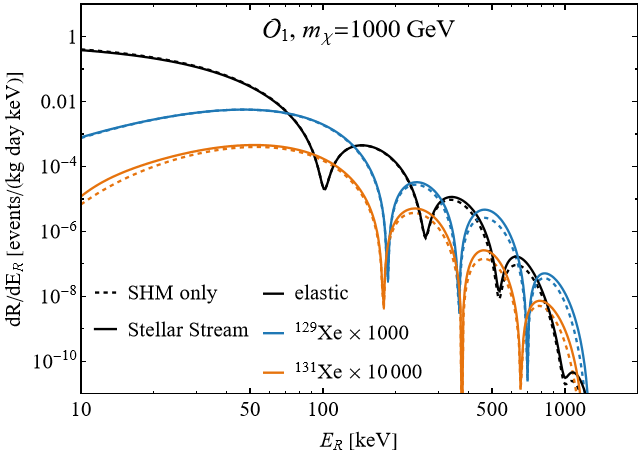}
    \includegraphics[
        width=0.48\linewidth,
        height=0.24\textheight,
        keepaspectratio
    ]{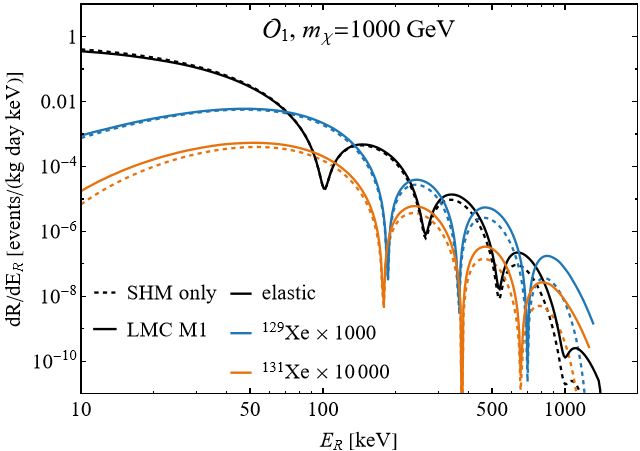}\\[0.5em]
    \includegraphics[
        width=0.48\linewidth,
        height=0.24\textheight,
        keepaspectratio
    ]{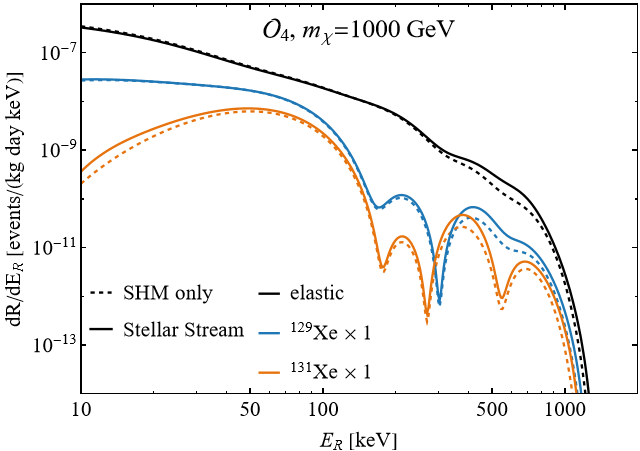}
    \includegraphics[
        width=0.48\linewidth,
        height=0.24\textheight,
        keepaspectratio
    ]{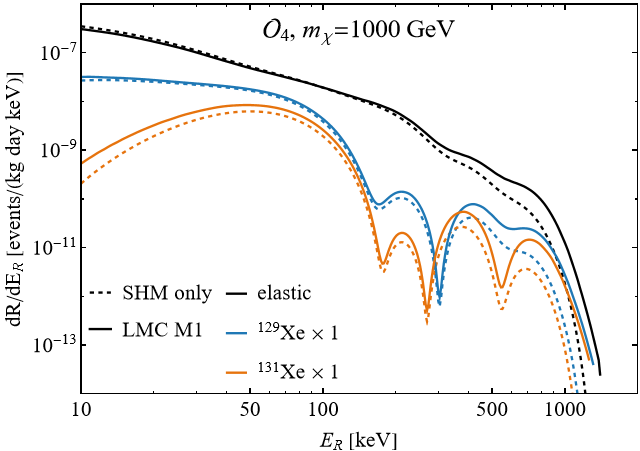}

    \caption{Similar to Fig.~\ref{fig:dsigmadE_Acc_1},
    but showing the results for $\mathcal{O}_1$ and $\mathcal{O}_4$.}
    \label{fig:dsigmadE_Acc_2}

    \vspace{1em}

    \includegraphics[
        width=0.48\linewidth,
        height=0.24\textheight,
        keepaspectratio
    ]{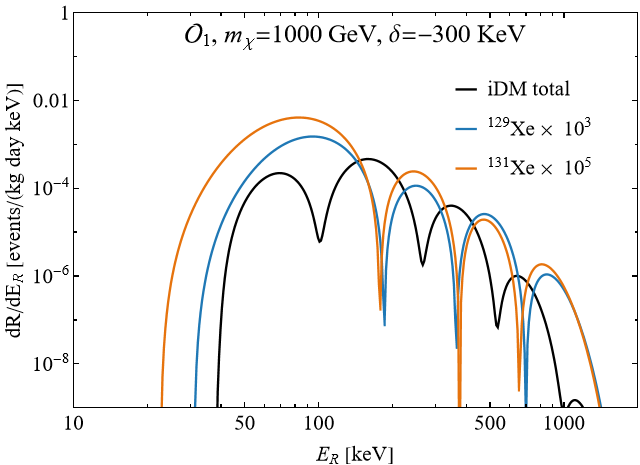}
    \includegraphics[
        width=0.48\linewidth,
        height=0.24\textheight,
        keepaspectratio
    ]{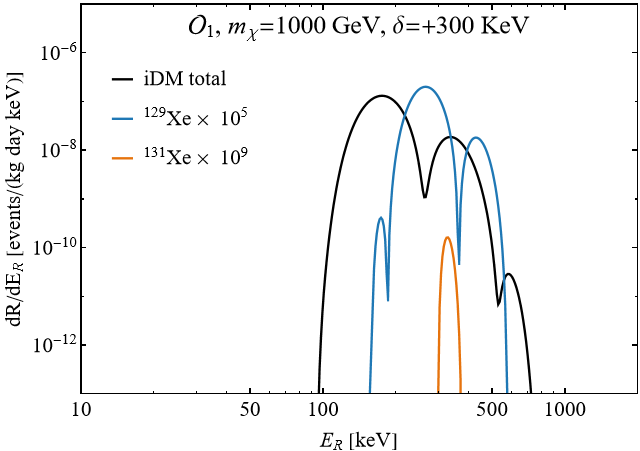}\\[0.5em]
    \includegraphics[
        width=0.48\linewidth,
        height=0.24\textheight,
        keepaspectratio
    ]{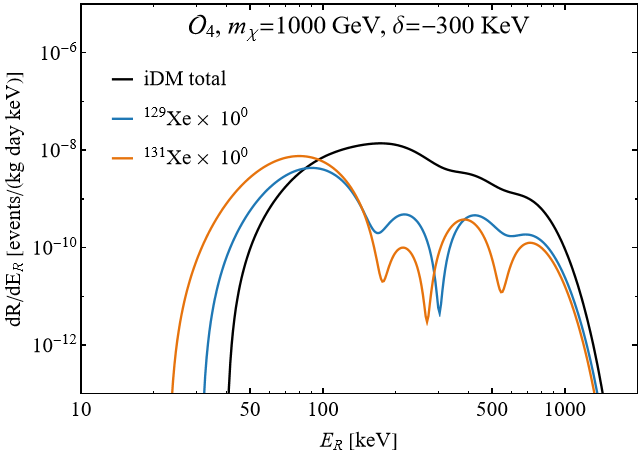}
    \includegraphics[
        width=0.48\linewidth,
        height=0.24\textheight,
        keepaspectratio
    ]{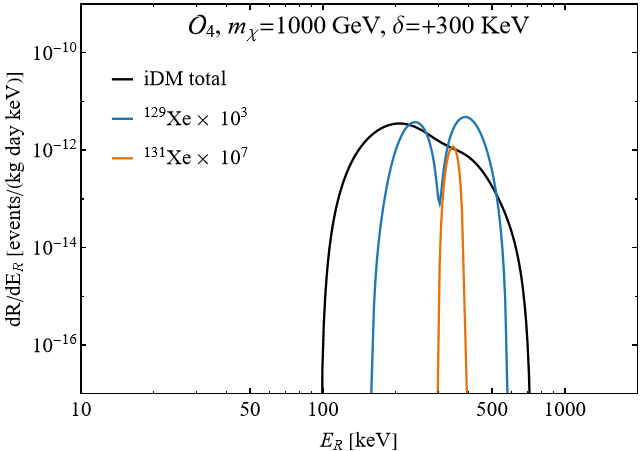}

    \caption{Similar to Fig.~\ref{fig:dsigmadE_iDM_1},
    but showing the results for $\mathcal{O}_1$ and $\mathcal{O}_4$.}
    \label{fig:dsigmadE_iDM_2}
\end{figure}

\clearpage
\section{Statistics of Alternate Inelastic-Xenon Signal Responses}
\label{app:electronvsphoton}
In this section, we list the counterparts of Table~\ref{tab:stat_WIMP}-\ref{tab:stat_iDM} with an alternative \texttt{NEST} setup, where the ${\rm Xe}^\ast$ de-excitation energy takes the form of an energetic electron instead of a photon.

\begin{table}[h!]
\centering
\small
\renewcommand{\arraystretch}{1.12}
\setlength{\tabcolsep}{3pt}
\begin{tabular}{c cc cc cc}
\toprule
& \multicolumn{2}{c}{$\mathcal O_{4}$} & \multicolumn{2}{c}{$\mathcal O_{6}$} & \multicolumn{2}{c}{$\mathcal O_{10}$} \\
\cmidrule(lr){2-3}
\cmidrule(lr){4-5}
\cmidrule(lr){6-7}
$m_\chi$ [GeV] & $N_{\mathrm{NR}}^{\mathrm{ref}}=1$ & $\Lambda=3\,\mathrm{TeV}$ & $N_{\mathrm{NR}}^{\mathrm{ref}}=1$ & $\Lambda=3\,\mathrm{TeV}$ & $N_{\mathrm{NR}}^{\mathrm{ref}}=1$ & $\Lambda=3\,\mathrm{TeV}$ \\
\midrule
\multicolumn{7}{c}{LMC} \\
\midrule
$90$ & $<0.01$ & $<0.01$ & $<0.01$ & $<0.01$ & $<0.01$ & $<0.01$ \\
$1000$ & $0.026$ & $<0.01$ & $0.012$ & $<0.01$ & $0.016$ & $<0.01$ \\
\midrule
\multicolumn{7}{c}{$90\%$ SHM $+\,10\%$ Stellar Stream} \\
\midrule
$90$ & $<0.01$ & $<0.01$ & $<0.01$ & $<0.01$ & $<0.01$ & $<0.01$ \\
$1000$ & $0.022$ & $<0.01$ & $0.012$ & $<0.01$ & $0.013$ & $<0.01$ \\
\bottomrule
\end{tabular}
\caption{Similar to Table~\ref{tab:stat_LMC_S1}, but using the pure electron response instead of the photon response for the nuclear de-excitation energy.}
\label{tab:stat_LMC_S1_beta}
\end{table}

\begin{table}[h!]
\centering
\small
\renewcommand{\arraystretch}{1.12}
\setlength{\tabcolsep}{3pt}
\begin{tabular}{cc cc cc cc}
\toprule
& & \multicolumn{2}{c}{$\mathcal O_{4}$} & \multicolumn{2}{c}{$\mathcal O_{6}$} & \multicolumn{2}{c}{$\mathcal O_{10}$} \\
\cmidrule(lr){3-4}
\cmidrule(lr){5-6}
\cmidrule(lr){7-8}
$m_\chi$ [GeV] & $\delta$ [keV] & $N_{\mathrm{NR}}^{\mathrm{ref}}=1$ & $\Lambda=3\,\mathrm{TeV}$ & $N_{\mathrm{NR}}^{\mathrm{ref}}=1$ & $\Lambda=3\,\mathrm{TeV}$ & $N_{\mathrm{NR}}^{\mathrm{ref}}=1$ & $\Lambda=3\,\mathrm{TeV}$ \\
\midrule
\multirow{7}{*}{$90$} & $-300$ & $0.11$ & $0.046$ & $0.041$ & $<0.01$ & $0.051$ & $<0.01$ \\
 & $-200$ & $0.051$ & $0.028$ & $0.035$ & $<0.01$ & $0.037$ & $<0.01$ \\
 & $-50$ & $<0.01$ & $<0.01$ & $<0.01$ & $<0.01$ & $<0.01$ & $<0.01$ \\
 & $0$ & $<0.01$ & $<0.01$ & $<0.01$ & $<0.01$ & $<0.01$ & $<0.01$ \\
 & $+50$ & $<0.01$ & $<0.01$ & $<0.01$ & $<0.01$ & $<0.01$ & $<0.01$ \\
 & $+200$ & $<0.01$ & $<0.01$ & $<0.01$ & $<0.01$ & $<0.01$ & $<0.01$ \\
 & $+300$ & $<0.01$ & $<0.01$ & $<0.01$ & $<0.01$ & $<0.01$ & $<0.01$ \\
\midrule
\multirow{7}{*}{$1000$} & $-300$ & $0.17$ & $<0.01$ & $<0.01$ & $<0.01$ & $0.024$ & $<0.01$ \\
 & $-200$ & $0.13$ & $<0.01$ & $0.014$ & $<0.01$ & $0.032$ & $<0.01$ \\
 & $-50$ & $0.032$ & $<0.01$ & $0.015$ & $<0.01$ & $0.019$ & $<0.01$ \\
 & $0$ & $0.019$ & $<0.01$ & $0.012$ & $<0.01$ & $0.013$ & $<0.01$ \\
 & $+50$ & $0.016$ & $<0.01$ & $<0.01$ & $<0.01$ & $<0.01$ & $<0.01$ \\
 & $+200$ & $0.035$ & $<0.01$ & $<0.01$ & $<0.01$ & $<0.01$ & $<0.01$ \\
 & $+300$ & $<0.01$ & $<0.01$ & $<0.01$ & $<0.01$ & $<0.01$ & $<0.01$ \\
\bottomrule
\end{tabular}
\caption{Similar to Table~\ref{tab:stat_iDM}, but using the $\beta$-electron response for the nuclear de-excitation energy and including the $\delta=0$ WIMP benchmarks corresponding to Table~\ref{tab:stat_WIMP}.}
\label{tab:stat_iDM_beta}
\end{table}

\end{document}